\documentclass[12pt]{iopart}
\usepackage{iopams}
\input epsf
\def\be{\begin{equation}}
\def\ee{\end{equation}}
\def\bs{\begin{subequations}}
\def\es{\end{subequations}}
\def\ben{\begin{enumerate}}
\def\een{\end{enumerate}}
\def\ba{\begin{eqnarray}}
\def\ea{\end{eqnarray}}

\def\al{\alpha}

\def\bx#1{{\bf #1}}
\def\cx#1{{\cal #1}}
\def\tx#1{{\tilde{#1}}}
\def\hx#1{{\hat{#1}}}

\def\nin{\noindent}

\def\us#1{\underline{#1}}
\def\hth/#1#2#3#4#5#6#7{{\tt hep-th/#1#2#3#4#5#6#7}}
\def\atmp#1({Adv.\Theor.\Math.\Phys.\ $\us {#1}$\(}
\def\nup#1({Nucl.\ Phys.\ $\us {B#1}$\ (}
\def\plt#1({Phys.\ Lett.\ $\us  {B#1}$\ (}
\def\cmp#1({Comm.\ Math.\ Phys.\ $\us  {#1}$\ (}
\def\prp#1({Phys.\ Rep.\ $\us  {#1}$\ (}
\def\prl#1({Phys.\ Rev.\ Lett.\ $\us  {#1}$\ (}
\def\prv#1({Phys.\ Rev.\ $\us  {#1}$\ (}
\def\mpl#1({Mod.\ Phys.\ Let.\ $\us  {A#1}$\ (}
\def\atmp#1({Adv.\ Theor.\ Math.\ Phys.\ $\us  {#1}$\ (}
\def\ijmp#1({Int.\ J.\ Mod.\ Phys.\ $\us{A#1}$\ (}

\def\dwn{\Delta^*_{W_{n+1}}}

\def\zh{\hat{z}}
\def\lra{\leftrightarrow}

\def\dwn{\Delta^*_{W_{n+1}}}

\def\hx#1{{\hat{#1}}}

\def\br{\hfill\break}

\def\foot#1{\footnote{#1}}

\def\lra{\leftrightarrow}
\def\IP{\bx P}

\newcommand{\req}[1]{(\ref{#1})}

\def\subsubsection#1{\noindent \ \\ {\it #1}\vskip 0.10cm}

\begin{document}
\jl{6}
\begin{flushright}
 CERN--TH/99-327\\
hep-th/nnnmmyy
\end{flushright}

\title[Non-perturbative $\cx N=1$ strings from geometric singularities]
{Non-perturbative $\cx N=1$ strings from geometric \\ singularities}

\author{P. Mayr}

\address{TH Division, CERN, 1211 Geneva 23, Switzerland }

\def\Large{\large}
\def\LARGE{\large\bf}

\vskip 1cm

\def\cx#1{{\cal #1}}
\begin{abstract}
The study of curved D-brane geometries in type II strings 
implies a general relation between local singularities $\cx W$ of 
Calabi--Yau manifolds and gravity free supersymmetric 
QFT's. The minimal supersymmetric case is described by F-theory 
compactifications on $\cx W$ and can be used as a starting point to
define minimal supersymmetric 
heterotic string compactifications on compact Calabi--Yau manifolds
with holomorphic, stable gauge backgrounds. The geometric 
construction generalizes to non-perturbative 
vacua with five-branes and provides a framework to study non-perturbative 
dynamics of the heterotic theory.
\end{abstract}
\vskip 6cm

\noindent
{\tt To appear in the proceedings of the} STRINGS '99  {\tt conference,
Potsdam (Germany), 19--25 July 1999}
\vskip 0.5cm

\noindent
CERN--TH/99-327

\maketitle\vfil\break

\section{Introduction}
Much of the work in string theory in the last years has been
devoted to dualities between seemingly different physical 
theories. A remarkable aspect of these dualities is that 
they involve often a geometrization of physics, that is there 
is a correspondence of physical quantities with 
data of a geometric object. In extreme cases, one arrives at
a duality between a physical theory and a geometric theory,
with the physical data such as the moduli space, symmetries,
spectrum and even correlation functions being in one-to one
correspondence with the quantities of a geometric theory. In particular,
if the coupling constants of the physical theory are mapped
to geometric parameters, the duality provides an extremely powerful 
tool to study quantum effects in the physical theory.

Generally, the most useful formulation of the geometric theory 
is in terms of complex geometries $W$, with the complex deformations
of $W$ being equivalent to the moduli space $\cx M_{phys}$ 
of scalar vev's $a_i$ of the
physics theory. The geometry $W$ can be defined as the zero locus 
of a holomorphic polynomial in some ambient space $Y$
\be
W:\ p_W(x_i,u_\al)=0,
\ee
where $x_i$ are holomorphic coordinates on $Y$ and $u_\al$ parametrize
the complex structure of $W$. The idea is that there is a map
$u_\al \to a_i$ which maps the complex structure moduli space of 
$W$ to $\cx M_{phys}$. The prototype for the relation between
a complex geometry $W$ and a physical system is the case of 
$\cx N=2$ $SU(2)$ Super-Yang-Mills theory in four-dimensions, whose
one-dimensional moduli space on the Coulomb branch has been shown
to be equivalent to the complex deformations
of a torus \cite{sw}. However it was realized shortly later \cite{kklmv,
klmvw} that 
this relation is only a special example of a very general connection
between complex geometries of Calabi--Yau 3-fold singularities
and $\cx N=2$ supersymmetric gravity free quantum field theories
in four dimensions. The link between
geometry and physics is provided
by D-brane geometries in type IIB strings in the following way%
\footnote{For a review and more references, see \cite{Lec}.}.
If the type IIB string is compactified on a Calabi--Yau 3-fold $X$
with small 3-cycles $C_i$,
there are extra light states 
from D3-branes wrapped on the $C_i$. In the limit of vanishing volume
these states give rise to massless charged particles 
in uncompactified space-time \cite{stro},
with their quantum numbers and interactions determined by the geometry
of the cycle $C_i$ embedded in $X$. Since the type II string coupling
is in the hypermultiplet sector, which does not couple neutrally to the
vector multiplets, the exact effective two derivative action
of the vector multiplets on the Coulomb branch is given by the tree level
type IIB physics, which in turn is determined by the
classical complex geometry. 
In the limit of very small 3-cycles, which means we consider singular
geometries, one can decouple gravity and most of the fundamental
string spectrum and obtains an equivalence between the complex deformations
of Calabi--Yau 3-fold
singularities and the Coulomb branch of 
general gravity free $\cx N=2$ QFT's in four dimensions.
Apart from the exact non-perturbative effective action, there
are other important non-perturbative
data of the field theory, such as the stable BPS spectrum \cite{klmvw}
and S-duality symmetries \cite{kmv}, 
that are determined in the exact tree level
type IIB theory and thus, the geometry of the singularity.

This relation between geometric Calabi--Yau 3-fold singularities and
gravity free QFT's can be extended \cite{BM}\footnote{See ref.
\cite{Lec2} for an introduction to the subject.} 
to theories with $\cx N=1$ 
supersymmetries by replacing the type II string theory with 
F-theory \cite{vafaf}. In this case we obtain field theories that are
embedded in minimal supersymmetric string theories in $d<8$ dimensions,
in particular  $\cx N=1$ supersymmetric string theories in four dimensions.

An especially interesting class of the latter
consists of heterotic string compactifications
on Calabi--Yau 3-folds $Z$, which are the most promising perturbative string
theories from the phenomenological point of view. However an aspect of
these so-called $(0,2)$ vacua that has turned out to be difficult to study,
is the specification of a suitable gauge background, which 
takes the form of a holomorphic stable vector bundle (or sheaf) on $Z$.
It turns out \cite{BM} that these bundles appear as the moduli space of 
a certain class of field theories that are "dual" to Calabi--Yau 
4-fold singularities in the previous sense and can be constructed and studied
using the geometric duality, very much as in the case of $\cx N=2$
supersymmetric QFT's. Again we obtain non-perturbative information
from the geometric perspective and in particular the geometric 
construction generalizes to  non-perturbative vacua of the heterotic
string, including those with background 5-branes \cite{DMW} 
and non-perturbative gauge symmetries. 

\section{Singularities for $\cx N=1$ supersymmetric theories}
We have mentioned already that the crucial link between the geometric
singularities and the physical system associated to it is provided by
the curved geometry of D-branes of type II strings wrapped on vanishing
cycles. However, to reduce supersymmetry, we need to compactify 
F-theory on the singularity, rather than type II strings. In general,
F-theory has a very different way to generate massless states,
in particular  from open strings stretching between background D7 branes.
However note that in the duality between F-theory compactified on a Calabi--Yau
$n$-fold $W_n$ times a torus $T^2$ and type IIA on $W_n$, the radii of
$T^2$ are mapped to geometric moduli of the type IIA theory on $W_n$.
We can therefore  
describe the minimal supersymmetric theory from F-theory on $W_n$ in
terms of a type IIA compactification on the F-theory-limit
of $W_n$, defined as the patch in the geometric moduli of type IIA on $W_n$ 
that corresponds to the decompactification of the $T^2$ factor 
in the dual F-theory compactification.
As a matter of fact this allows to adopt many of the methods 
developed in the $\cx N=2$ context to the minimal supersymmetric 
case. In particular one can establish
in this way the following geometric duality:
\begin{table}[h]
\scriptsize
\centerline{\vbox{\tabskip=0pt\offinterlineskip
\def\tablerule{\noalign{\hrule}}
\def\gap{\omit&height4pt&&&&\cr}
\halign{\strut#&\vrule#\tabskip=1em plus2em&
\hfil~#~\hfil&\vrule#&
\hfil~#~\hfil&\vrule#
\tabskip=0pt\cr
\tablerule\gap
&& Calabi--Yau Geometry $\cx W_{n+1}$
&&$\cx N=1$ Field theory  $(Z_n,V)$&\cr
\gap\tablerule\gap
&& \hbox{Local singularities $\cx W_{n+1}$ 
} && Compactification of $\cx N=1$, 10d SYM&\cr
&&\hbox{in 
$n+1$-fold fibrations of} && \hbox{on ell. fibered Calabi--Yau $Z_n$;}&\cr
&& \hbox{elliptically fibered ALE spaces} 
&& \hbox{Holomorphic stable vector bundle $V$ on $Z_n$}&\cr
\gap\tablerule
}}}
\end{table}

Similarly as in the $\cx N=2$ context, one needs 
only type IIA (and F-theory) physics to establish the above
equivalence. A necessary assumption that is
needed is that $Z_n$ is elliptically fibered. The 
equivalence holds for any structure group $H$ of the bundle.

Note that the moduli space of the $\cx N=1$ field
theory appearing on the right hand side of 
the above table is precisely the same as that 
of the gauge background of a heterotic compactification on $Z_n$ in 
the point particle limit, under the condition
that the structure $H$ fits in the perturbative heterotic gauge symmetry 
$G_0^{het}$, {\it e.g.} $E_8\times E_8$.
Since one arrives at the above equivalence
using only type IIA/F-theory physics, one can
{\it derive} type IIA/heterotic and F-theory/heterotic duality 
in the point particle limit. Note that the equivalence is true for {\it any}
$H$; for $H\subset G^{het}_0$ it becomes equivalent to a 
duality with the heterotic string in the point particle limit.

Since the field theory involved in the geometric duality is so closely related to a 
string theory compactification, about which we do not have a good control, 
it makes sense to consider the reverse process of decoupling gravity and find {\it global}
embeddings of the singularities $\cx W_{n+1}$ 
into compact Calabi--Yau manifolds $W_{n+1}$ to describe
dual pairs of the full F-theory/heterotic and IIA/heterotic dualities away from the 
point particle limit.
Interestingly one finds that global embeddings exist 
precisely if $H\in G^{het}_0$. While it is
expected from the physics point of view that we recover the known consistent string theories
in this way, the result is non-trivial from the geometric point of view.
The global embeddings define dual pairs of string duality (as opposed to
the relation between singularity and field theory), namely
a compact Calabi--Yau $n+1$-fold
$W_{n+1}$ on which F-theory is compactified and a compact Calabi--Yau 
$Z_n$ together with a family of holomorphic stable vector bundles $V$ on it that describe
the dual heterotic data.
Moreover one obtains 
a precise map of the moduli spaces of the two theories in the point particle
limit. We can then {\it use} the duality $W_{n+1} \leftrightarrow (Z,V)$ away from this limit
to study non-perturbative properties of the minimal supersymmetric heterotic string vacua.

\section{Vector bundles and type II strings}
At first sight it is not obvious that F-theory compactified 
on the geometric singularities $\cx W_{n+1}$ 
describes holomorphic 
stable vector bundles on compact Calabi--Yau manifolds  $Z_n$ of
one complex dimension less. The relation between the geometric singularity 
and vector bundles can be traced back to a symmetry of type IIA on K3$\times
T^2$ \cite{kmv} that is closely related to mirror symmetry on K3.
Consider a type IIA string compactified on a K3 surface. 
This gives an $\cx N=2 $ supersymmetric theory in six dimensions.
From dimensional reduction  we obtain twenty 
$\cx N=2$ vector multiplets which contain  as their bosonic degrees 
of freedom each one vector and four real scalars. The latter
transform as a $\us{4}$ under an $SO(4)_R$ R-symmetry and parametrize the 58 metric moduli of K3
plus 22 values of the $B$-fields on $H_2(K3)$.  

In a a given algebraic realization $M_2$ of K3, 
the 80 moduli of K3 split
in complexified K\"ahler and complex structure moduli\footnote{Subscripts,
as in $M_2$,
denote the complex dimension of a geometry.}. The
$SO(4)_R$ rotations acting on the scalars in $\us{4}$ do not
preserve this split in general. In particular there is an $SO(4)_R$
rotation that exchanges K\"ahler and complex structure moduli spaces
$\cx M_{KM}$ and $\cx M_{CS}$, respectively. Such a transformation
is usually known as {\it mirror symmetry} of K3 manifolds, which 
says that a type IIA string compactified on $M_2$ describes the same
physics as a type IIA string compactification on a "different"\foot{Since
the K3 surface is unique, mirror symmetry acts as a discrete 
identification in the moduli space of K3 surfaces.} K3
manifold $W_2$, with the K\"ahler and complex structure moduli exchanged.
The new algebraic 
K3 surface $W_2$ is called the mirror manifold of $M_2$. We have the
following identifications under mirror symmetry:
\def\cxms{\cx M^{string}_E}
\be\label{msi}
\qquad \qquad \qquad \qquad \qquad M_2 \lra W_2,
\ee
$$
\cx M_{CS}(M_2) \lra \cx M_{KM}(W_2),\qquad
\cx M_{KM}(M_2) \lra \cx M_{CS}(W_2) .\nonumber
$$

\nin
There is an intriguing generalization of this symmetry upon 
compactification on a further $T^2$ to four dimensions. 
In this case we obtain an $\cx N=4$ supersymmetric string theory.
The $\cx N=4$ vector multiplet now contains 6 scalars 
which transform in a $\us 6$ of an $SO(6)_R$ symmetry.
The two extra scalars are the internal components $A_i$ of the six-dimensional
gauge fields on the $T^2$, call it $E$,
and thus parametrize the stringy moduli space of
Wilson lines on the elliptic curve, $\cxms(H)$. Here $H$ denotes
the structure group of the Wilson line background. Again the 
$SO(6)_R$ rotations provide identifications within the type IIA
moduli space $\cx M_{IIA}(K3\times T^2)$:
\be\label{msii}
\cx M_{KM}(M_2) \lra \cx M_{CS}(W_2) \lra \cxms(H).
\ee
In particular there is now an element of $SO(6)_R$, corresponding
to the last arrow, which provides a
relation between
the moduli space of stringy
Wilson lines on $E$ with the K\"ahler moduli space of an algebraic
K3 manifold. 

\begin{figure}[h]
\hbox to\hsize{\hss
\epsfysize=3.5cm
\epsffile{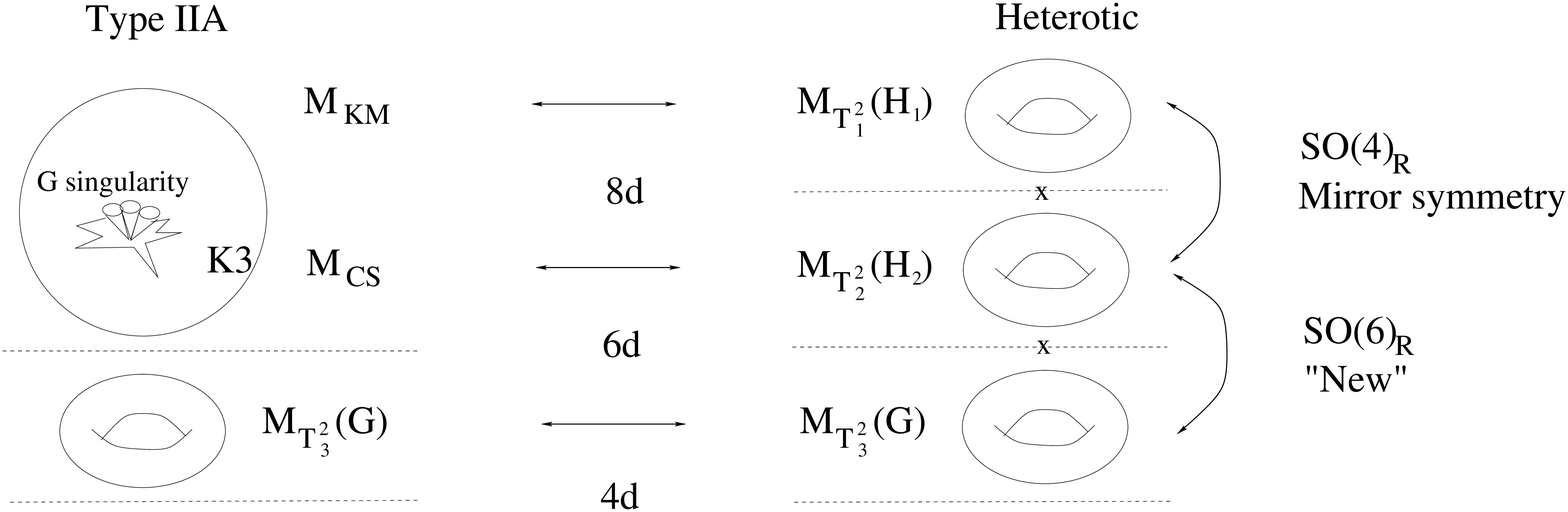}\hss}
\caption{The symmetry \req{msii} from a heterotic point of view.}
\label{f:fhetdualii}\end{figure}

Let us illustrate the symmetry \req{msii} of the type IIA compactification on 
K3 $\times T^2$ in the dual heterotic picture where it is very natural;
however we stress that we do not need the heterotic dual to derive the
symmetry. Fig. \ref{f:fhetdualii} shows the type IIA compactification
and the corresponding dual heterotic string compactified on 
$T_1^2\times T_2^2\times T_3^2$. 
\vspace{0cm}
The factorization of $T^6$ on the heterotic side is related to
$i)$ the elliptic fibration and $ii)$ the factorization of the gauge 
background on K3 $\times T^2$ in the type IIA theory. Under these circumstances, the K\"ahler moduli of K3 can be identified with moduli space $\cx M_{T^2_1}(H_1)$
of $H_1$ Wilson lines on the first $T^2_1$ and 
similarly complex structure moduli
with Wilson lines in the second factor $\cx M_{T^2_2}(H_2)$.
The non-abelian gauge symmetry in six dimensions is $G$ which on the
type IIA side arises from the $G$ singularity and on the heterotic side is
the commutant of $H_1\times H_2$ in $G^{het}_0$. Upon further compactification
to four dimensions on a torus $T^2_3$, we obtain another factor 
$\cx M_{T^2_3}(G)$ on both sides.

The heterotic theory has an obvious symmetry under permutation of the 
three $T^2$ factors. The exchange of the first two tori amounts to
mirror symmetry in the type IIA theory. On the other hand an exchange
that involves the third torus gives a new symmetry of the moduli
space of the type IIA compactification that identifies Wilson lines
with structure group $G$ on $T^2$ with either K\"ahler or complex
deformations of a $G$ singularity of K3. 
This is the new relation described in \req{msii}.

Finally we can decouple most of the fundamental string states and 
gravity in the limit of a singular geometry
of the K3, or in terms of physics, in the field theory limit. 
The elliptically fibered 
2-fold singularities $\cx M_2$ that describe the local 
patch in $M_2$ (and similarly a local geometry $\cx W_2$ for 
the complex geometry side) obtained in this limit describe 
field theory Wilson lines on a torus $E$:
\be\label{e:equi}
M_{KM}(\cx M_2) \cong  M_{CS}(\cx W_2) \cong M^{FT}_{E}(H).
\ee
Note that this relation will be more general as the one obtained 
in the global K3 context. In the latter, $r={\rm rank}\ H$ 
is bounded from above by the dimension of $H_2(K3)$, whereas in 
\req{e:equi} we can actually consider local singularities of ALE spaces
of arbitrary rank. 

The above field theory limit is compatible with taking the F-theory limit
of the type IIA theory. The F-theory compactification on $\cx W_2$ 
gives then the deliberate geometric dual we are looking for.

\section{Holomorphic stable bundles on Calabi---Yau $n$-folds}
In the previous section we have seen how holomorphic (semi-)stable 
bundles on a Calabi--Yau 1-fold, or simply Wilson lines on $T^2$,
are related to F-theory compactified on a 2 complex dimensional
Calabi--Yau singularity $\cx W_2$. To obtain holomorphic stable bundles on
Calabi--Yau $n$-folds, we consider holomorphic 
fibrations of $\cx W_2$ over some $n-1$ complex dimensional base $B_{n-1}$.
The total space $\cx W_{n+1}$, which is required to satisfy 
the Calabi--Yau condition in order to be
a valid type IIA and F-theory background, is an $n+1$ dimensional
non-compact 
singularity with an elliptic fibration inherited from that of $\cx W_2$.

The dual heterotic picture is the following \cite{FMW}: 
the data of $H$ Wilson lines on $T^2$ are
fibered over the same base $B_{n-1}$, in virtue of the adiabatic principle 
\cite{VWi}. The $T^2$ fibration over $B_{n-1}$ results in a 
compact Calabi--Yau $n$-fold $Z_n$ on which the heterotic string is 
compactified and the gauge background on $T^2$ extends to a family of 
holomorphic
stable bundles on $Z_n$.

Let us now sketch the general form of the dual singularities $\cx W_{n+1}$
which have a nice geometric and group theoretical interpretation.
Let $(y,x,z;v)$ denote a special set of coordinates on the elliptically
fibered ALE space $\cx W_2$; 
in particular $v$ is a coordinate on the base $\IP^1$
of the elliptic fibration of $\cx W_2$. Moreover the 
$x_i$ are coordinates on the base $B_{n-1}$ of the $\cx W_2$ fibration of 
$\cx W_{n+1}$.
The singularity $\cx W_{n+1}$ is defined as the vanishing of
the polynomial $p$:
\begin{equation}
\label{e:gfi}
p=p_0+p_+=
p_0(y,x,z|x_i)+\sum_{i} v^ip^i_+(y,x,z|x_i)=0.
\end{equation}
We have divided the polynomial $p$ in a $v$ independent
part $p_0$ and a $v$ dependent piece $p_+$. They describe the 
geometry $Z_n$ and the gauge bundle $V$ on $Z_n$, respectively.
In fact $p_0$ is the defining equation for the elliptically 
fibered, compact Calabi--Yau $Z_n$ in Weierstrass form. The polynomial
$p_+$ that describes the bundle has a very nice group theoretical 
structure. If we consider the extended Dynkin diagram of the 
structure group $H$ of $V$, then each node of index $s_i$ contributes
precisely one monomial $p_+^{s_i}$ to $p_+$. E.g., for $SU(N)$
we have $N$ nodes of index one and the equation for $p_+$ becomes
(for $N$ even)
\be
p_+=v^1\cdot(z^N\ a_N(x_i)+ xz^{N-2}\ a_{N-2}(x_i)+\dots+x^{N/2} a_0(x_i)).
\ee
Since in this case $v$ appears only linearly in $p$, it can be integrated
out as far as variation of the complex structure is concerned. This
yields the two separate conditions $p_0=0$ and $p^1_+$=0. The first 
equation defines $Z_n$. The $N$ roots of the second equation determine
$N$ points on the elliptic fiber $E$ of $Z_n$, with their position 
on $E$ varying with the values of the base variables $x_i$. We thus
recover the spectral cover construction of $SU(N)$ bundles on elliptic
fibrations described in $\cite{FMW,BJPS}$.

More generally, we obtain a similar geometric representation of
families of holomorphic stable bundles on $Z_n$ for any $H$. 
The actual construction of the geometries $\cx W_{n+1}$ can 
be phrased best in the framework of toric geometry. This approach has the
advantage of being very general, in particular in treating 
general Calabi--Yau manifolds $Z_n$ with non-smooth elliptic fibrations,
as well as studying singularities in both, the geometric and
gauge data. Moreover one can define a toric map that identifies
the dual objects 
\be
\qquad \qquad f:\ \cx W_{n+1}\leftrightarrow (Z_n,V),
\ee
with important physical data of the heterotic side, 
such as the geometry of the manifold $Z_n$, 
the family of bundles on it, non-perturbative 5-branes or gauge
symmetries and the stability of the bundle having a very simple and
systematic representation in terms of the toric geometry of the singularity
$\cx W_{n+1}$ on the left side \cite{BM,BMii}. This allows for a simple
construction of deliberate families of 
stable bundles $V$ with arbitrary structure
group $H$ on any toric Calabi--Yau in terms of the singularity $\cx W_{n+1}$.

Moreover, if $H\in G_0^{het}$ one can easily determine the possible
global embeddings of the local singularity $\cx W_{n+1}
\hookrightarrow W_{n+1}$, where $W_{n+1}$  is now the global
Calabi--Yau $n+1$-fold which gives the dual {\it string} theory
when used as an F-theory compactification. The map
$W_{n+1}\leftrightarrow (Z_n,V)$ can then be used to study non-perturbative 
properties of the minimal supersymmetric heterotic string theory on $Z_n$,
as is illustrated in the next section.

\section{Some results on (non-perturbative) $\cx N=1$ heterotic strings
\label{ss:nphetstring}}
Let us finally sketch some results on (non-perturbative) heterotic
physics which have been 
obtained in \cite{BM,BMii}\ using the above method. 

\subsubsection{Stability of the bundle $V$}
We complained already that it is very hard to find
explicit solutions to the stability equations for $V$ and in
the sequel we claimed that we can construct 
a large class of such stable bundles 
seemingly without effort using the geometric singularities $\cx W_{n+1}$. 
Obviously,
there has to be quite a good reason for this simplification in the 
geometric construction. Indeed there is one important property of the
geometric singularities $\cx W_{n+1}$ that we did not interpret so far in the 
heterotic duals: the Calabi--Yau condition that ensures that the type
IIA, or F-theory, compactified on $\cx W_{n+1}$ is consistent. Indeed
one can show \cite{BMii} that the Calabi--Yau property of $\cx W_{n+1}$
translates to a stability condition on the background $V$ on $Z_n$,
expressed in terms of a bound on the 
first Chern class $\eta=c_1(\cx N)$ of a 
line bundle $\cx N$ that is an important characteristic data of 
the vector bundle $V$ on $Z$ \cite{FMW}. One finds that 
\be\label{e:ebound}
\nu(G)\  c_1(\cx L) \leq \ \eta \ (\leq 12 c_1(\cx L)\ ),
\ee
where $G$ is the singularity type of the fiber geometry $W_2$ and
$\nu(G)$ is a certain characteristic number defined in \cite{BMii}.
Explicit expressions for $\cx N$ and the bounds on $\eta$
for toric bases, such as $\IP^2$ and $\bx F_n$
and blow ups thereof, can be found in 
\cite{BM}\foot{A bound on $\eta$ derived in \cite{RA} is
different from those in \cite{BM} and \cite{BMii}.}.

The answer to why it was so easy to get stable solutions from the
geometric construction is thus that we have replaced the complicated 
condition for stability in the heterotic string, about which we have poor
control,  with the simple Calabi--Yau condition on the dual 
type IIA/F-theory side, which is much easier to deal with.

\subsubsection{Standard embedding}
A simple solution to solve the background equations 
for $V$ is to set the
gauge connection equal to the spin connection of the manifold,
$V=TZ$. This is the so-called standard embedding. 
Though this configuration is not too appealing for
phenomenology - in four dimensions,
the heterotic gauge group is $E_6\times E_8$ and
the matter spectrum is tied to the Euler number of $Z$ - it 
has been studied extensively in the past because it was one
of the few known solutions. The structure group of $V$ is 
$SU(n)$ for a Calabi--Yau $n$-fold.

The global geometry $W_{n+1}$ corresponding to the F-theory dual 
can be determined in the following way. From the fact
that the bundle in the second factor is trivial it follows that 
the line bundle $\cx N$ 
is trivial 
in the second $E_8$ factor. The Weierstrass form for $W_{n+1}$ and $Z_n$ 
takes the following form:
\ba\label{tbiii}
p_{Z_n}&&=y^2+x^3+x\tx z^4f+\tx z^6g,\cr
p_{W_n}&&=y^2+x^3+x(\zh z w)^4f+(\zh z w)^6g+\zh^6z^5w^7\Delta+\zh^6z^7w^5.
\ea
Here $(z,w)$ denote the variables parametrizing the base $\IP^1$
of the elliptically fibered K3 fiber $W_2$ of the 
fibration $\pi_F:W_{n+1}\to B_{n-1}$ and 
$\Delta=4f^3+27g^2$
is the discriminant of the elliptic fibration of $Z_n$.
This generalizes the six-dimensional result 
obtained in \cite{AD} to four and lower dimensions.

\subsubsection{Non-perturbative gauge symmetries}
In the F-theory picture, gauge symmetries
correspond to singularities in the elliptic fibration. Those located
over the base $B_{n-1}$ have a perturbative interpretation in the
heterotic dual, whereas non-perturbative ones are located over 
curves in $B_{n-1}$ \cite{MV}. Using the map
$W_{n+1}\to (Z_n,V)$ we can describe the heterotic compactification 
that leads to the same dynamics non-perturbatively. In the six-dimensional 
case we find:
\vskip0.2cm

\leftskip .5cm \rightskip .5cm
\nin {\it  $(*)$
Consider the $E_8\times E_8$ string compactified on 
an elliptically fibered K3  with a singularity of type 
$G$ at a point $s=0$ and a special gauge background $\hx V$. If
the restriction of the spectral cover\footnote{For $G\neq SU(N)$ we use
the generalization of the spectral cover in terms of 
the sections of a certain weighted
projective bundle over $B_{n-1}$ \cite{FMW}.} 
of $\hx V_{|E}$ to the fiber $E$ at $s=0$ is
sufficiently trivial, the heterotic string acquires a 
non-perturbative gauge symmetry $G_{np} \supset G$. }
\vskip0.2cm

\leftskip .0cm \rightskip .0cm
\nin
The triviality 
condition can be made precise by specifying
the behavior of $V$ near $s=0$ \cite{BM}. Similar results hold
for four-dimensional compactifications.

We point out that the above triviality condition on the spectral cover
does not imply that the field strength of $V$, which measures the behavior
of $V$ near the singularity,
is trivial.
In fact it has been shown recently in \cite{witk3} that if $F=0$
on the singularity, the conformal field theory of the heterotic
string is well behaved and there is no non-perturbative gauge
symmetry.

\subsubsection{Non-perturbative dualities}
The map $f:\ W_{n+1} \leftrightarrow (Z_n,V)$ can be ambiguous in the sense
that there are two (or even more) ways to associate a pair
$(Z_n,V)$ to $W_{n+1}$. If the two maps are compatible with
the same elliptic fibration, we obtain a non-perturbative duality
of two heterotic string theories
\be\label{duaii}
(Z_n,V)\sim (Z_n^\prime,V^\prime).
\ee
The conditions under which
this duality exists, can be formulated in simple properties of
the toric polyhedron $\dwn$ associated to the Calabi--Yau 
$W_{n+1}$ in its construction as a toric hypersurface \cite{BM}.
The generic form of the duality is the following:
\vskip0.2cm

\leftskip 0.5cm \rightskip 0.5cm\nin 
{\it $(**)$ Let the heterotic string be compactified  
on a Calabi--Yau three-fold with $G^\prime$ singularity and 
with a certain gauge background with structure group  $H$
such that the toric data $\dwn$ fulfil the above mentioned condition.
Then there exists a 
non-perturbatively equivalent compactification
on a Calabi--Yau manifold 
with $G$ singularity and with a specific gauge background
with structure group  $H^\prime$}. 
\vskip0.2cm

\leftskip 0.cm \rightskip 0.cm\nin 
Here $H$ $(H^\prime)$
is the commutant of $G$ $(G^\prime)$ in $E_8\times E_8$. Note that
the duality exchanges the groups of the geometric singularity and the
gauge bundle: the singularity in the dual manifold is the commutant of
the structure group and {\it vice versa}.
Let
us give an extreme example of such an duality: 
the heterotic string with a trivial gauge bundle on a smooth
K3 has a perturbative $E_8\times E_8$ gauge symmetry and
in addition $n_T^\prime=24$ non-perturbative tensor multiplets
from the 24 five-branes required to satisfy anomaly cancellation
in six dimensions. The dual theory
is a heterotic theory with $E_8\times E_8$ gauge bundle on 
a K3 with $E_8\times E_8$ singularity. The perturbative
gauge symmetry is trivial, while non-perturbative dynamics
associated to the singularity produce both, the $E_8\times E_8$
gauge symmetry as well as 24 extra tensor multiplets.

This duality reminds very much of a known duality in the linear sigma model
formulation of heterotic strings,
namely a symmetry under the exchange
of the data defining the manifold and the data defining the bundle
\cite{DK}.
However note that in our case this duality is 
in general {\it non-perturbative}.
 
\subsubsection{Mirror symmetry of F-theory}
Consider a six-dimensional F-theory compactification on $W_3$
and the associated heterotic dual $(Z_2,V)$ as defined above.
If the mirror manifold $M_3$ of $W_3$ is also elliptically
fibered, there is also a heterotic theory $(Z_2^\prime,V^\prime)$
associated to it and one can ask the question of how the two theories 
are related. The answer is that after compactification on a 
three-torus to three dimensions, the two become equivalent in virtue
of mirror symmetry of type II strings 
\cite{IS}.

A proposal for the relation between $(Z_2,V)$ and $(Z_2^\prime,V^\prime)$
has been made in \cite{PR}
based on a comparison of hodge numbers 
and gauge symmetries in some cases. Using the toric map
$W_{n+1}\to (Z_n,V)$ one can determine the two theories; in fact one finds
a subtle realization of Higgs and Coulomb branches in toric geometry
as expected from the action of three-dimensional mirror symmetry \cite{BM}.
The general relation is of a similar form as in $(**)$ above:
a heterotic theory with a $G$ bundle compactified on a K3 manifold
with $\hx H$ singularity gets mapped to a heterotic theory 
with a bundle with structure group $\hx H$ compactified on a
manifold with $G$ singularity. Note that we have an explicit
map of the moduli spaces of the two theories. It would be interesting
to formulate the associated three-dimensional dual pairs
associated to this relation, as in 
\cite{HOV}.
\vskip 2cm

\section{Outlook}
The geometric construction of minimal supersymmetric
heterotic string vacua in four dimensions and their F-theory duals
opens an interesting way to study 
phenomenologically relevant string compactifications. 
The identifications between $W_{n+1}$ and $(Z_n,V)$ may serve as a 
starting point to study more refined non-perturbative relations 
between the dual F- and heterotic string theory. In particular we
would like to formulate both perturbative and non-perturbative 
quantities of the heterotic string, e.g. gauge and Yukawa couplings as
well as superpotential of the $(0,2)$ vacua, in terms of geometric 
quantities on $W_{n+1}$. A concrete example are the
correlation functions determined by holomorphic prepotentials,
considered in \cite{4fs}\footnote{See also \cite{LS} for an interesting
study of correlation functions in 
the eight-dimensional duality.}. Also for the superpotential, a 
geometrical approach exists and has been studied in \cite{KS,4fs,KVsp,Vsp,Dia}.
In the ideal (and supposedly too optimistic) case, 
we can hope to obtain exact results for at least some of the holomorphic 
physical quantities in the $\cx N=1$ theory from geometric information
on $W_{n+1}$, similarly as it happened to work in the 
case of $\cx N=2$ supersymmetry.
\\

\noindent
{\bf Acknowledgements}\br
I would like to thank Per Berglund for the collaboration on this work
and the organizers of Strings '99 for the
opportunity to present it and for their hospitality.

\end{document}